\documentstyle[amssymb,aps,12pt]{revtex}

\begin{document}
\title{Dynamics of a two-component atomic Bose-Einstein condensate }
\author{Ping Zhang$^1$, C.K. Chan$^2$, Xiang-Gui Li$^3$, Xian-Geng Zhao$^1$}
\address{$^1$Institute of Applied Physics and Computational Mathematics, Beijing\\
100088, China \\
$^2$Department of Applied Mathematics, The Hong Kong Polytechnic University,%
\\
Kowloon, Hong Kong\\
$^3$Department of Applied Mathematics, University of Petroneum, Shandong\\
257062, China}
\maketitle

\begin{abstract}
The dynamical population oscillations between two internal states of a
Bose-Einstein condensate are investigated within the rotating wave
approximation. Analytical expressions for the population imbalance in the
number states and coherent states have been derived, which predict different
revival periods. Thus the true quantum state of the condensates may be
unambiguously determined by detecting the atom intensity evolution for one
internal state.%
\newline%
\newline%
\end{abstract}

The experimental observation of the Bose-Einstein condensation of a trapped,
dilute gas of alkali atoms, and the high accuracy of the engineering, are
opening a new avenue to investigate the interplay between macroscopics and
quantum coherence. In particular, recent experiments on two-component
Bose-Einstein condensates (BECs) in $^{87}$Rb\cite{Andrews,Hall} have
stimulated considerable theoretical work on the dynamics of the phase and
number fluctuations of Bose condensates. Vigorous efforts have concentrated
on the issue of temporal phase coherence between two coupled Bose
condensates. Along semiclassical (mean) analysis much attention has focused
on the coherent phenomena such as Josephson effect and macroscopic quantum
self-trapping (MQST)\cite{ref2,ref3,ref4,ref5}. Studies involving small
quantum corrections show that due to the nonlinearities arising from the
atom-atom interactions and the discrete spectrum of the many-body system,
the dynamics of the population oscillations arising in the mean-field
approximation is modulated by the collapses and revivals\cite
{ref2,ref6,ref7,ref8,ref9,ref10,ref11,ref14}.

In the present work, we study quantum dynamics of a two-component atomic BEC
such as $^{87}$Rb. Our purpose is to provide an analysis on how the coherent
atom oscillations between two components depend on the initial states chosen
for the condensates. We chose two different initial states of the
condensate. In the first example we consider Fock state for each component
of the condensate. In this case, we show that the revival period depends on
whether the total atom number $N$ is even or odd. In the other example we
invoke Bose-broken symmetry and consider the product of two coherent states.
In this case we show that the revival period is always definite. In
experiments, such as $^{87}$Rb, initial Fock states can be created by
condensing independently two thermal atomic clouds trapped in different
Zeeman level. Whereas, the initial coherent state may be created by
splitting a condensate (trapped, say, in the Zeeman state $%
|f=2,m_{f}=2\rangle $) in two condensates trapped in the Zeeman states $%
|f=2,m_{f}=2\rangle $ and $|f=2,m_{f}=1\rangle $, respectively. Another
aspect we are interested in this paper is the dependence of the coherent
atom oscillations on the relative phase between two internal states and
nonlinear atomic collisions.

We consider here Bose-Einstein condensation of a trapped gas of atoms that
have two internal states $|1\rangle $ and $|2\rangle $ with energies $%
\hslash \omega _o/2$ and $-\hslash \omega _o/2$, respectively. There is a
spatially uniform radiation field with frequency $\omega _e$ that couples
the two internal states with a Rabi frequency $\Omega _0$. The atoms in
states $|1\rangle $ and $|2\rangle $ are subject to isotropic harmonic
trapping potentials $V_i({\bf r})=\frac 12m\omega _ir^2$ for $i=\{1,2\}$,
respectively. Furthermore, the atoms interact via elastic two-body
collisions through the interaction potentials $V_{ij}({\bf r}-{\bf r}%
^{\prime })=\frac{4\pi \hslash ^2a_{ij}}m\delta ({\bf r}-{\bf r}^{\prime })$%
, Where $a_{ij}$ is the $s$-wave scattering length between atoms in states $%
i $ and $j$. It is assumed that $a_{ij}>0$ corresponding to repulsive
interactions. In the formalism of second quantization, the system is
described by the Hamiltonian

\begin{equation}
\widehat{H}=\widehat{H}_{atom}+\widehat{H}_{coll},  \eqnum{1a}
\end{equation}
\begin{eqnarray}
\widehat{H}_{coll} &=&\int d^3r\{\frac{4\pi \hslash ^2a_{11}}{2m}\widehat{%
\Psi }_1^{+}({\bf r})\widehat{\Psi }_1^{+}({\bf r})\widehat{\Psi }_1({\bf r})%
\widehat{\Psi }_1({\bf r})  \nonumber \\
&&+\frac{4\pi \hslash ^2a_{22}}{2m}\widehat{\Psi }_2^{+}({\bf r})\widehat{%
\Psi }_2^{+}({\bf r})\widehat{\Psi }_2({\bf r})\widehat{\Psi }_2({\bf r}) 
\nonumber \\
&&+\frac{4\pi \hslash ^2a_{12}}m\widehat{\Psi }_1^{+}({\bf r})\widehat{\Psi }%
_2^{+}({\bf r})\widehat{\Psi }_1({\bf r})\widehat{\Psi }_2({\bf r})\}, 
\eqnum{1b}
\end{eqnarray}
\begin{eqnarray}
\widehat{H}_{atom} &=&\int d^3r\{\widehat{\Psi }_1^{+}({\bf r})\left[ -\frac{%
\hslash ^2}{2m}\nabla ^2+V_1({\bf r})+\frac{\hslash \delta }2\right] 
\widehat{\Psi }_1({\bf r})  \nonumber \\
&&+\widehat{\Psi }_2^{+}({\bf r})\left[ -\frac{\hslash ^2}{2m}\nabla ^2+V_2(%
{\bf r})-\frac{\hslash \delta }2\right] \widehat{\Psi }_2({\bf r})  \nonumber
\\
&&+\frac{\hslash \Omega _0}2\left( \widehat{\Psi }_1^{+}({\bf r})\widehat{%
\Psi }_2({\bf r})+\widehat{\Psi }_2^{+}({\bf r})\widehat{\Psi }_1({\bf r}%
)\right) \}.  \eqnum{1c}
\end{eqnarray}
Here, the atomic field operators $\widehat{\Psi }_i({\bf r})$ and $\widehat{%
\Psi }_i^{+}({\bf r})$ have been written in the field interaction
representation which is rotating at the frequency of the external field $%
\omega _e$, $\delta =\omega _0-\omega _e$ denotes the field detuning from
resonance excitation.

Now we can approximate the field operators in the zero-temperature limit by
a two-mode model such that $\widehat{\Psi }_i({\bf r})\approx \widehat{c}%
_i\phi _i({\bf r})$, where $\widehat{c}_i$ is the annihilation operator
which obeys the usual boson commutation relations. In the two-mode
approximation the Hamiltonian becomes (with $\hslash =1$),

\begin{eqnarray}
\widehat{H} &=&(\frac{1}{2}\delta +E_{1})\widehat{c}_{1}^{+}\widehat{c}%
_{1}+(-\frac{1}{2}\delta +E_{2})\widehat{c}_{2}^{+}\widehat{c}_{2}+\frac{%
\Omega }{2}(\widehat{c}_{1}^{+}\widehat{c}_{2}+\widehat{c}_{2}^{+}\widehat{c}%
_{1})  \nonumber \\
&&+\frac{{\sl \kappa }_{11}}{2}\widehat{c}_{1}^{+}{}^{2}\widehat{c}%
_{1}{}^{2}+\frac{{\sl \kappa }_{22}}{2}\widehat{c}_{2}^{+}{}^{2}\widehat{c}%
_{2}{}^{2}+\kappa _{12}\widehat{c}_{1}^{+}\widehat{c}_{1}\widehat{c}_{2}^{+}%
\widehat{c}_{2},  \eqnum{2}
\end{eqnarray}
where

\begin{equation}
E_{i}=\int d^{3}r\phi _{i}^{\ast }({\bf r})[-\nabla ^{2}/(2m)+V_{i}({\bf r}%
)]\phi _{i}({\bf r}),  \eqnum{3a}
\end{equation}
\begin{equation}
\Omega =\Omega _{0}\int d^{3}r\phi _{2}^{\ast }({\bf r})\phi _{1}({\bf r}), 
\eqnum{3b}
\end{equation}
\begin{equation}
\kappa _{ij}=\frac{4\pi a_{ij}}{m}\int d^{3}r|\phi _{i}({\bf r})|^{2}|\phi
_{j}({\bf r})|^{2}.  \eqnum{3c}
\end{equation}

The analysis of Eq. (2) is greatly simplified by the introduction of the
angular momentum operators

\begin{equation}
\widehat{J}_x=\frac 12(\widehat{c}_2^{+}\widehat{c}_2-\widehat{c}_1^{+}%
\widehat{c}_1),  \eqnum{4a}
\end{equation}

\begin{equation}
\widehat{J}_y=\frac i2(\widehat{c}_2^{+}\widehat{c}_1-\widehat{c}_1^{+}%
\widehat{c}_2),  \eqnum{4b}
\end{equation}

\begin{equation}
\widehat{J}_z=\frac 12(\widehat{c}_1^{+}\widehat{c}_2+\widehat{c}_2^{+}%
\widehat{c}_1).  \eqnum{4c}
\end{equation}
The Casimir invariant is therefore

\begin{equation}
\widehat{J}^{2}=\frac{\widehat{N}}{2}(\frac{\widehat{N}}{2}+1),  \eqnum{5}
\end{equation}
where $\widehat{N}=\widehat{c}_{1}^{+}\widehat{c}_{1}+\widehat{c}_{2}^{+}%
\widehat{c}_{2}$ is the total number operator, which is a constant of motion
and thus can be set equal to the total number of atoms $N$. Therefore the
present two-mode approximation is transformed into an angular momentum model
with total angular momentum given by $j=N/2$. In terms of angular momentum
operators the Hamiltonian (2) may be rewritten as 
\begin{equation}
\widehat{H}=\Omega \widehat{J}_{z}+2\kappa \widehat{J}_{x}^{2}+f(j), 
\eqnum{6}
\end{equation}
where

\begin{equation}
\kappa =\frac{1}{2}(\frac{\kappa _{11}}{2}+\frac{\kappa _{22}}{2}-\kappa
_{12}),  \eqnum{7a}
\end{equation}
\begin{equation}
\kappa ^{\prime }=\frac{1}{2}(\frac{\kappa _{11}}{2}+\frac{\kappa _{22}}{2}%
+\kappa _{12}),  \eqnum{7b}
\end{equation}
\begin{equation}
f(j)=2\kappa ^{\prime }j^{2}+(E_{1}+E_{2}-\kappa -\kappa ^{\prime })j. 
\eqnum{7c}
\end{equation}
In deriving Eq. (6) we have assumed that 
\begin{equation}
E_{1}+\delta -E_{2}+(\frac{\kappa _{11}}{2}-\frac{\kappa _{22}}{2})(2j-1)=0,
\eqnum{8}
\end{equation}
a condition that can always be achieved by adjusting field detuning $\delta $
and shifting the energy levels of the condensate components. Note that some
control of the nonlinear parameter $\kappa $ can be achieved through the
proper engineering of the trapping potential or tuning the scattering
lengths via Feshbach resonances.

In general the Hamiltonian (6) is difficult to treat in an exact way because
of the presence of nonlinear atomic interactions. In order to gain physical
insight into the dynamics of such a two-component BEC problem, some
approximations are necessary: a common assumption\cite{ref15,Kuang2} is the
so-called rotating wave approximation ({\bf RWA}) $\widehat{U}=e^{-i\Omega t%
\widehat{J}_z}$. Using the operator identity 
\begin{equation}
\exp (i\lambda \widehat{J}_z)\widehat{J}_x\exp (-i\lambda \widehat{J}_z)=%
\widehat{J}_x\cos \lambda -\widehat{J}_y\sin \lambda ,  \eqnum{9}
\end{equation}
the Hamiltonian in the rotating frame ({\bf RF}) becomes 
\begin{eqnarray}
\widehat{H}_{RF} &=&\widehat{U}^{+}\widehat{H}\widehat{U}+i\frac{d\widehat{U}%
^{+}}{dt}\widehat{U}  \eqnum{10} \\
&=&\frac 12\kappa (e^{2i\Omega t}\widehat{J}_{+}^2+e^{-2i\Omega t}\widehat{J}%
_{-}^2+\widehat{J}_{+}\widehat{J}_{-}+\widehat{J}_{-}\widehat{J}_{+})+f(j), 
\nonumber
\end{eqnarray}
where $\widehat{J}_{\pm }=\widehat{J}_x\pm i\widehat{J}_y$ are raising and
lowing operators. Now we invoke {\bf RWA} by suppressing rapidly oscillating
time-dependent terms in Eq. (10). Thus we obtain the {\bf RWA }Hamiltonian
in the rotating frame as follows 
\begin{equation}
\widehat{H}_{RF}^{(r)}=-\kappa \widehat{J}_z^2+g(j),  \eqnum{11}
\end{equation}
where $g(j)=\kappa j^2+f(j)$. Instead of discussing dynamics in the rotating
frame, we return to the Hamiltonian in the previous field interaction
representation

\begin{equation}
\widehat{H}_{eff}=\widehat{U}\widehat{H}_{RF}^{(r)}\widehat{U}^{+}-i\frac{d%
\widehat{U}^{+}}{dt}\widehat{U}=-\kappa \widehat{J}_z^2+\Omega \widehat{J}%
_z+g(j),  \eqnum{12}
\end{equation}
which is our starting Hamiltonian for the following discussions. It reveals
in Eq. (12) that $\widehat{J}_z$ is now a constant of motion and solely
determined by the initial condition. The frozen $\widehat{J}_z$ behavior can
be understood when $\Omega $ is sufficiently larger than $\kappa $, i.e., $%
\Omega \gg \kappa $. In this case, the external field forces the total spin
to remain polarized in the $z$-direction because it costs energy to change
the spin direction. To see the validity of our {\bf RWA} Hamiltonian, we
numerically integrate the Sch\"{o}dinger equation with the original
Hamiltonian (6) and present in Fig. 1(a) the expectation values $\langle 
\widehat{J}_z\rangle $ against time for various values of $\Omega $ with the
initial angular momentum state $|j,-j\rangle _x$ where the subscript denotes
the quantization axis along the $x$ direction. We remark that for the system
starts from $|j,-j\rangle _x$, the only nonvanishing spin component is $%
\widehat{J}_z$ because $\langle $ $\widehat{J}_x\rangle =\langle $ $\widehat{%
J}_y\rangle =0$ at all times. Clearly it reveals in Fig. 1(a) that when $%
\Omega =0$, $\langle \widehat{J}_z\rangle $ vanishes after some time, which
means that the initial coherence and polarization of $\widehat{J}_z$ is
completely destroyed by the phase diffusion due to nonlinear term $2\kappa 
\widehat{J}_x^2$. However, when $\Omega $ is sufficiently larger than $%
\kappa $, the value of $\langle \widehat{J}_z\rangle $ remains almost
unchanged during time evolution, consistent with Eq. (12) in which $\widehat{%
J}_z$ is a constant of motion. As a further illustration we calculate the
energy spectrums of the two Hamiltonians and present the results in Fig.
1(b). It is found in calculations that the difference between the energies
of two Hamiltonians approaches to be negligible when increasing the value of 
$\Omega $. Thus we expect that the {\bf RWA} Hamiltonian is valid in the
parameter regime $\Omega \gg j\kappa $. In the following we will neglect the
constant term $g(j)$ because this term denotes a simple constant energy
shift and has no contribution to the dynamics of the system.

Now the Heisenberg equations of motion for the raising and lowing operators (%
$\widehat{J}_{\pm }=\widehat{J}_x\pm i\widehat{J}_y$ ) can be readily solved
as follows

\begin{equation}
\widehat{J}_{+}(t)=e^{it(-2\kappa \widehat{J}_z+\kappa +\Omega )}\widehat{J}%
_{+},  \eqnum{13a}
\end{equation}
\begin{equation}
\widehat{J}_{-}(t)=e^{it(2\kappa \widehat{J}_z+\kappa -\Omega )}\widehat{J}%
_{-}.  \eqnum{13b}
\end{equation}
Given the initial state $\mid \psi (0)\rangle $ of the system, a quantity
tailored to the dynamics is the population imbalance between the two
components

\begin{eqnarray}
N_{-}(t) &=&2\langle \psi (0)\mid \widehat{J}_x(t)\mid \psi (0)\rangle 
\eqnum{14} \\
&=&e^{i(\kappa +\Omega )t}\langle \psi (0)\mid e^{-2i\kappa t\widehat{J}_z}%
\widehat{J}_{+}\mid \psi (0)\rangle +c.c..  \nonumber
\end{eqnarray}

{\it Number state description. }We shall now proceed to demonstrate how the
population imbalance evolves for two different kinds of initial states of
the condensates. First we consider the case that the condensate is in the
Fock (number) states, equally the angular momentum states $\mid j,m\rangle _x
$ where the value of $2m$ gives the atom number difference between the two
BECs. We suppose that the condensates are initially localized in one well,
corresponding to the state $\mid j,j\rangle _x$. Substituting this initial
state into Eq. (14) gives

\begin{eqnarray}
N_{-}(t) &=&e^{i(\kappa +\Omega )t}\sum_{m=-j}^j\sqrt{(j+m)(j-m+1)} 
\nonumber \\
&&\times d_{j,m}^j(-\frac \pi 2)d_{m-1,j}^j(\frac \pi 2)e^{-i2\kappa
mt}+c.c.,  \eqnum{15}
\end{eqnarray}
where the matrix elements $d_{m,m^{\prime }}^j(\theta )=\langle j,m\mid \exp
(-i\theta \widehat{J}_y)\mid j,m^{\prime }\rangle $. Using the identity

\begin{equation}
d_{j,m}^{j}(\theta )=(-)^{j-m}\left( 
\begin{array}{c}
2j \\ 
j+m
\end{array}
\right) ^{1/2}\left( \cos \frac{\theta }{2}\right) ^{j+m}\left( \sin \frac{%
\theta }{2}\right) ^{j-m},  \eqnum{16}
\end{equation}
and recursive relation

\begin{equation}
d_{j,m}^{j}(\theta )=-\frac{\sin (\theta /2)}{\cos (\theta /2)}\left( \frac{%
j+m+1}{j-m}\right) ^{1/2}d_{j,m+1}^{j}(\theta ),  \eqnum{17}
\end{equation}
we obtain

\begin{eqnarray}
N_{-}(t) &=&\frac 1{2^N}e^{i(\kappa +\Omega )t}e^{i2j\kappa
t}\sum_{m=-j}^j(j+m)\left( 
\begin{array}{c}
2j \\ 
j+m
\end{array}
\right) e^{-i2\kappa (j+m)t}+c.c.  \nonumber \\
&=&N[\cos (\kappa t)]^{N-1}\cos (\Omega t).  \eqnum{18}
\end{eqnarray}

From Eq. (18) one can see that the population imbalance involves a rapidly
oscillating part and an envelope function that is responsible for the
collapse and revival of the population imbalance. The most prominent feature
in Eq. (18) is that the parity of total atom number $N$, whether it is odd
or even, plays an essential role in determining revival periods. When $N$ is
odd, the revival period is $\pi /\kappa $, and the population imbalance is
always larger than zero. That is, the population undergoes the macroscopic
quantum self-trapping. However, when the number of atoms is even, $N-1$ is
odd and the condensate will undergo antirevivals when $t=(2n+1)\pi /\kappa $
so that $N_{-}(t)=-N$ at these times. In this case, the revival period
becomes $2\pi /\kappa $ and the time average of $N_{-}(t)$ is zero. Two
examples are plotted in Figs. 2(a)-(b). To show the validity of the
analytical results of Eq. (18) which is derived from the {\bf RWA}
Hamiltonian, we numerically calculate the Schr\"{o}dinger equation with the
original Hamiltonian (6). The results are shown in Fig. 2(c)-(d),
corresponding to Fig. 2(a)-(c), respectively. Clearly, the {\bf RWA}
formula, equation (18), describes the system's evolution very well when
compared with the exact numerical results shown in Fig. 2(c)-(d), implying
the different periods of quantum revivals for different parities of $N$. We
notice that the reduction in revival amplitude shown in Fig. 2(c)-(d) is due
to the state mixing induced by angular momentum operator $\widehat{J}_x$,
which is eliminated in the RWA treatment. If the Fock states with a fixed
number of atoms are appropriate for the description of the condensates, then
by monitoring the initially unoccupied internal state using off-resonant
light scattering, this change for the magnitude of the revival period of the
population imbalance between the two components may be observed. A similar
phenomenon has been discussed in Ref.[16] in the context of the interference
pattern, without considering coupling and nonlinear atomic collision between
the two components. However, because the visibility of the interference
pattern is never negative, so the revival period considered in Ref.[16] is
always $\pi /\kappa $ with a $\pi $ phase shift for different parities of $N$%
.

{\it Coherent state description}. We turn now to the coherent state
description of the problem. It is customary to consider the condensates to
be in two coherent states, associated with a macroscopic wave function with
both magnitude and phase, the presence of which is due to the spontaneous
breaking of gauge symmetry. This is the view we will adopt here, to assume
the condensate to be such coherent states with definite relative phase

\begin{equation}
\mid \psi (0)\rangle =\mid \alpha ,\beta \rangle .  \eqnum{19}
\end{equation}
The amplitude of the two components may be presented by $\left| \alpha
\right| =\sqrt{N}\sin (\theta )$ and $\left| \beta \right| =\sqrt{N}\cos
(\theta )$, conserving total atom number. The relative phase between the two
components is defined to be $\phi $. Thus the two complex amplitudes are
related by $\beta \alpha ^{*}=\frac 12N\sin (2\theta )\exp (i\phi )$. To
give a closed expression for Eq. (14) with the initial state Eq. (19), we
need the normal-ordering arrangement for the operator function $\exp
(\lambda \widehat{J}_z)\widehat{J}_{+}$. With the aid of the following
identity

\begin{equation}
e^{-\lambda \widehat{c}_{i}^{+}\widehat{c}_{i}}=\sum_{r=0}^{\infty }\frac{%
x^{r}}{r!}\widehat{c}_{i}^{+r}\widehat{c}_{i}^{r},  \eqnum{20}
\end{equation}
where $x=e^{-\lambda }-1$, and after a straightforward calculation, we obtain

\begin{eqnarray}
e^{2\lambda \widehat{J}_z}\widehat{J}_{+} &=&\frac 12e^\lambda (\widehat{c}%
_2^{+}+\widehat{c}_1^{+})\sum_{m,n,p,q=-\infty }^\infty \{\frac{x^mx^ny^qz^p%
}{m!n!p!q!}  \nonumber \\
&&\times \widehat{c}_1^{+}{}^m\widehat{c}_1^{+n}\widehat{c}_2^{+p}\widehat{c}%
_2^{+q}[\sum_{l=-\infty }\frac{z^l}{l!}(\widehat{c}_2^{+}+\widehat{c}%
_1^{+})^l(\widehat{c}_2+\widehat{c}_1)^l]  \nonumber \\
&&\times \widehat{c}_1^m\widehat{c}_1^p\widehat{c}_2{}^n\widehat{c}_2^q\}(%
\widehat{c}_2-\widehat{c}_1),  \eqnum{21}
\end{eqnarray}
where $y=(e^{2\lambda }+1)/2$, and $z=(e^{2\lambda }-1)/2$. Substituting
Eqs. (20)-(21) into Eq. (14), we have the final result for the population
imbalance at time $t$ in the coherent state representation

\begin{equation}
N_{-}(t)=A\exp (-2N\sin ^2\frac{\kappa t}2)\cos [W\sin (\kappa t)+\Omega
t+\varphi ],  \eqnum{22}
\end{equation}
where we have defined $A=N\sqrt{1-\cos ^2\phi \sin ^2(2\theta )}$, $W=-N\sin
(2\theta )\cos \phi $, and a constant phase shift $\varphi =\arctan [\sin
\phi \tan (2\theta )]$. We can see from Eq. (22) that the dynamics of the
population imbalance is mainly determined by the envelope function $\exp
(-2N\sin ^2\frac{\kappa t}2)$, with modulated by a rapidly varying cosine
term. The characteristics of collapses and revivals can be clearly seen from
Eq. (22). For comparison with Eq. (18) derived in the number state space,
let us chose the value of $\theta =0$, corresponding the case that the
condensate is initially localized in one internal state. Then $N_{-}(t)$ in
Eq. (22) reduces to

\begin{equation}
N_{-}(t)=N\exp (-2N\sin ^2\frac{\kappa t}2)\cos (\Omega t).  \eqnum{23}
\end{equation}
Although no relative phase information occurs in Eq. (23), quite different
from number states, revivals for the condensate in the coherent states occur
at definite times $t=2\pi n/\kappa $ ($n$ is an integer ), independent of
the parities of the average particle number $N$. As illustrated in Ref.[16],
this difference of revival periods between the Fock state and coherent state
descriptions originates from the fact that the atom number difference
operator $\widehat{c}_2^{+}\widehat{c}_2-\widehat{c}_1^{+}\widehat{c}_1$ is
quantized in units of $2$ when the condensates are in Fock states and units
of $1$ for coherent states.

We emphasize here that for any realistic experimental set up, the exact atom
number is unknown and maybe has a large number fluctuation over many
different experimental runs. As a result, what we obtain in Fock states
should be wieghted by a factor $p(N)$ describing the probability the state $%
|j,j\rangle _x$ occurs at the begaining. According to quantum measurement
theory, however, the atom number in any given experiment would be well
defined. Thus in the Fock-state description, the occurence of antirivaval
phenomenon, which feaures the parity of the atom number, is robust for any
single experiment. On the other hand, the antirevival phenomenon never
occurs in the coherent-state description, which can be seen from Eq. (23).
This is because the coherent state is a superposition of the Fock states
with the weighting factor satisfying the Poisson number distrbution.
Consequently, in each single experiment quantum fluctuations will be present
in the coherent state description, which smears out the antirevivals. This
fact allows us the possibility to detect the true quantum state of the
two-component BEC system in realistic experiment.

In the case of the values of nonlinear interactions satisfying the following
condition

\begin{equation}
\kappa =\frac 12(\frac{\kappa _{11}}2+\frac{\kappa _{22}}2-\kappa
_{12})\equiv 0,  \eqnum{24}
\end{equation}
our RWA treatment of Hamiltonian (6) is exact and Eq. (22) reduces to
description of a simple Rabi oscillation with period $2\pi /\Omega $. This
form of atomic collective excitation has been observed in Ref.[2]. The
existence of Rabi oscillation also depends on the initial state conditions,
which is reflected by the fact that the oscillation amplitude $A$ in Eq.
(22) is determined by the initial population distribution and relative phase
between the two components. In particular when the value of $\theta =\pi /4$%
, corresponding to equal atom distribution between the two internal states,
and $\phi =n\pi $, we arrive at the conclusion that the oscillation is
completely suppressed and no quantum tunneling occurs during time evolution.
Note that the present conditions of zero oscillation is identical to the
fixed points discussed in Ref.[5], in which the semiclassical approximation
has been adopted. In general the value of effective interaction $\kappa $ is
not zero and it can be adjusted through the proper engineering of the
trapping potential, and hence of the condensate wave functions $\phi _i({\bf %
r})$. To illustrate the interplay between the weak nonlinearity and strong
Josephson-like coupling, we present in Fig. (3) $N_{-}(t)$ [Eq. (23)] for
several values of $\kappa $. It reveals in Fig. (3) that the presence of the
effective nonlinear interaction modulates the fringe visibility of the
internal Rabi oscillations as a manner of collapses and revivals, as has
been verified by JILA group\cite{Hall}. When $\kappa $ increases to large
values, the rapid phase diffusion caused by the nonlinear interaction
destroys the collective excitation of atoms, corresponding to the complete
smearing of the Rabi oscillations, as illustrated in Fig. 3(d).

To see the effect of the initial relative phase on the dynamics of the
system, we present in Fig. 4 the time evolution of $N_{-}(t)$ for four
different values of $\phi $. The value of $\theta $ in these figures is
chosen to be $\pi /4$, corresponding to the two condensates with equal
atoms. We can see in Fig. 4 that the oscillation amplitude is maximal for
the values $\phi =\pi /2$. Experimental observation of this change of
oscillation amplitudes may help us determine the relative phase between the
two condensates.

We have restricted our analysis to zero temperature. Our study of the effect
of initial state selection on the consequent dynamics of a two-component
condensate should be extended to finite temperatures, where dissipative
effects associated with the thermal cloud of noncondensate atoms must be
included. Further study of damping in this system is needed.

In summary, with the help of rotating wave approximation, we have derived
analytical expressions for the population imbalance between the two Bose
condensates in the number states and coherent states. The time evolution of
the coherent atom oscillations is shown to be quite different between these
two state descriptions. For the condensates in number states, the revival
periods of the oscillations are either $\pi /\kappa $ or $2\pi /\kappa $,
depending on whether the total atom number is odd or even. Whereas, for the
condensates in coherent states, the revival periods are always $2\pi /\kappa 
$. So the true quantum state of the condensates may be unambiguously
determined by detecting the atom intensity evolution for one trap.

{\Large Figure captions}

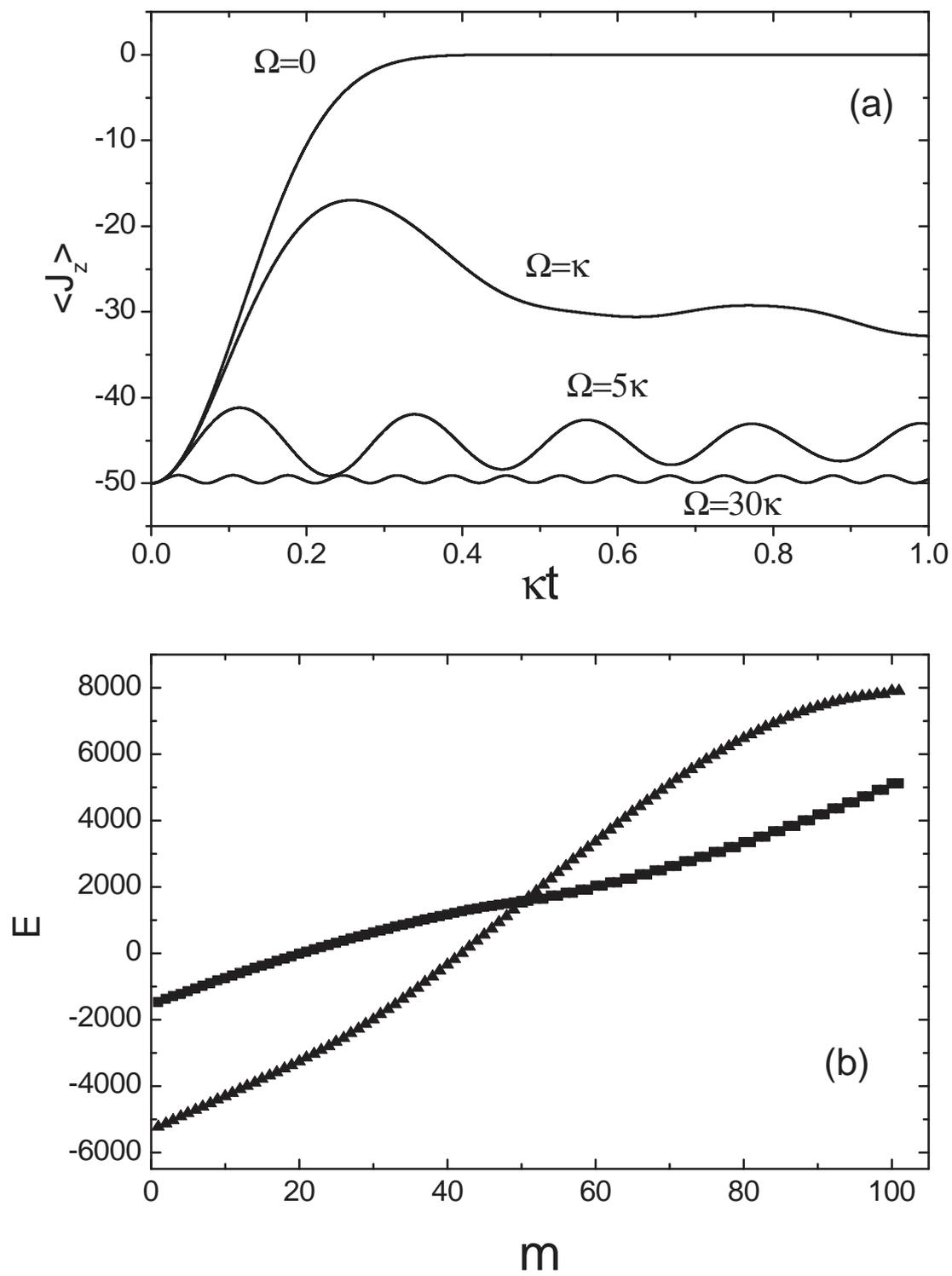
{\bf Fig. 1. }(a){\bf \ }Time evolution of $\langle \widehat{J}_z\rangle $
for the value of total atom number $N=100$. The initial state is $%
|j,-j\rangle _x$; (b) energy spectrums of the exact Hamiltonian (closed
triangles) and the {\bf RWA }Hamiltonian (closed squares) for the values of $%
N=100,\Omega =30\kappa $.

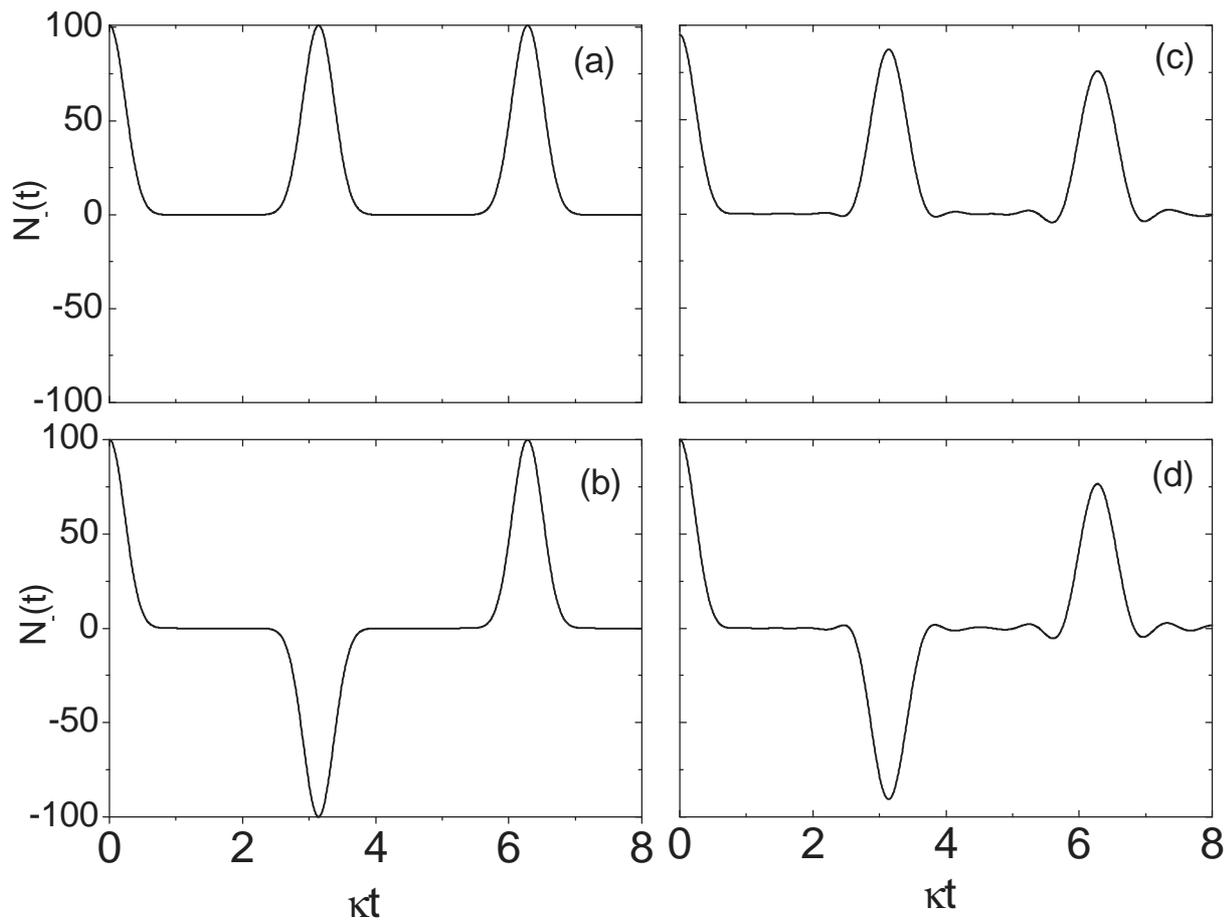
{\bf Fig. 2. }Analytical{\bf \ }results of{\bf \ }population imbalance $%
N_{-}(t)$ for the value of total atom number (a) $N=100$ and (b) $N=101$.
The exact numerical results corresponding to (a) and (b) are shown in (c)
and (d), respectively. Other parameters are $\kappa =0.001\Omega $.

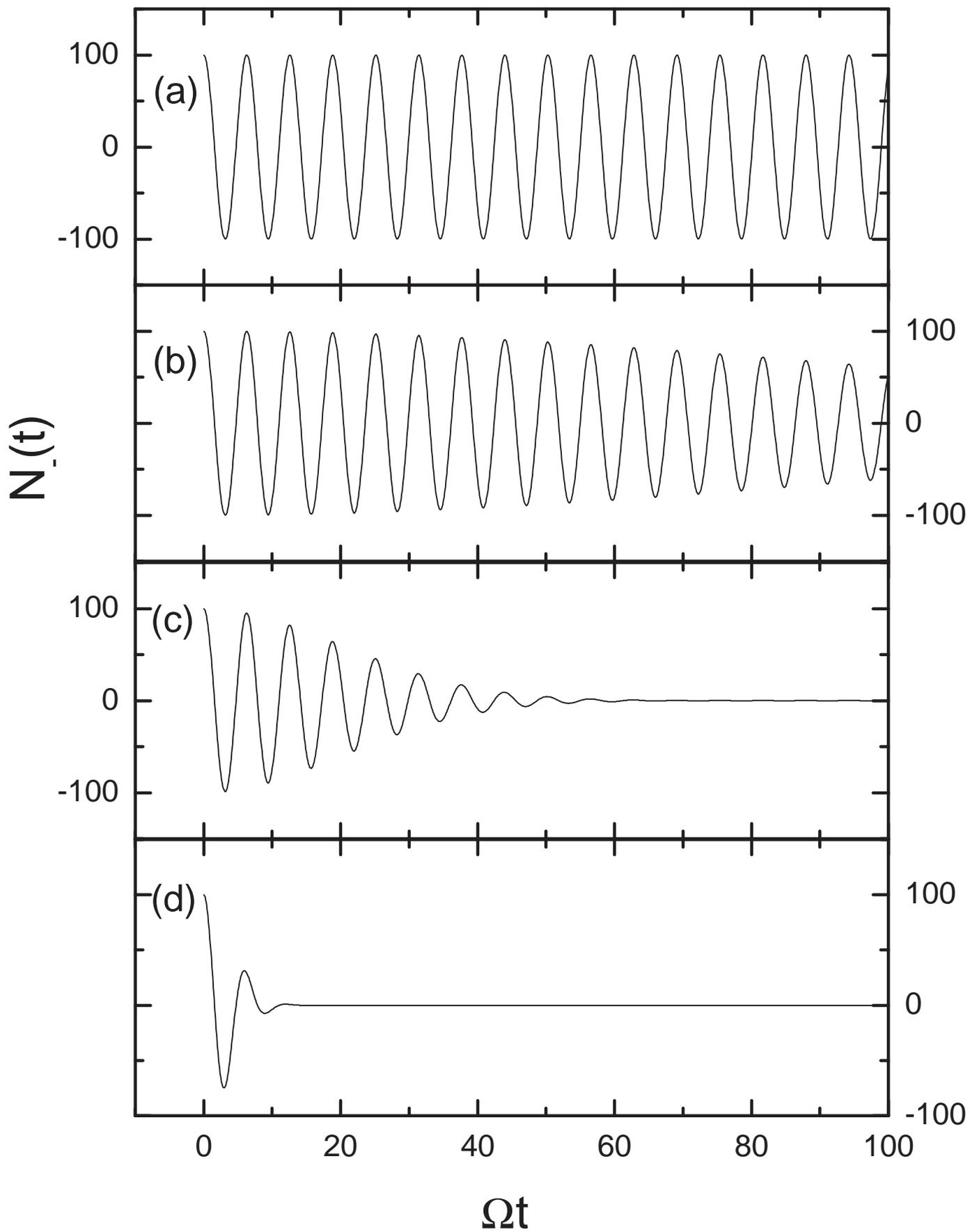
{\bf Fig. 3. }Population imbalance $N_{-}(t)$ for the value of effective
nonlinear coupling (a) $\kappa =0$, (b) $\kappa =0.001\Omega $, (c) $\kappa
=0.005\Omega $, and (d) $\kappa =0.025\Omega $. Other parameters are $N=100$
and $\theta =0$.

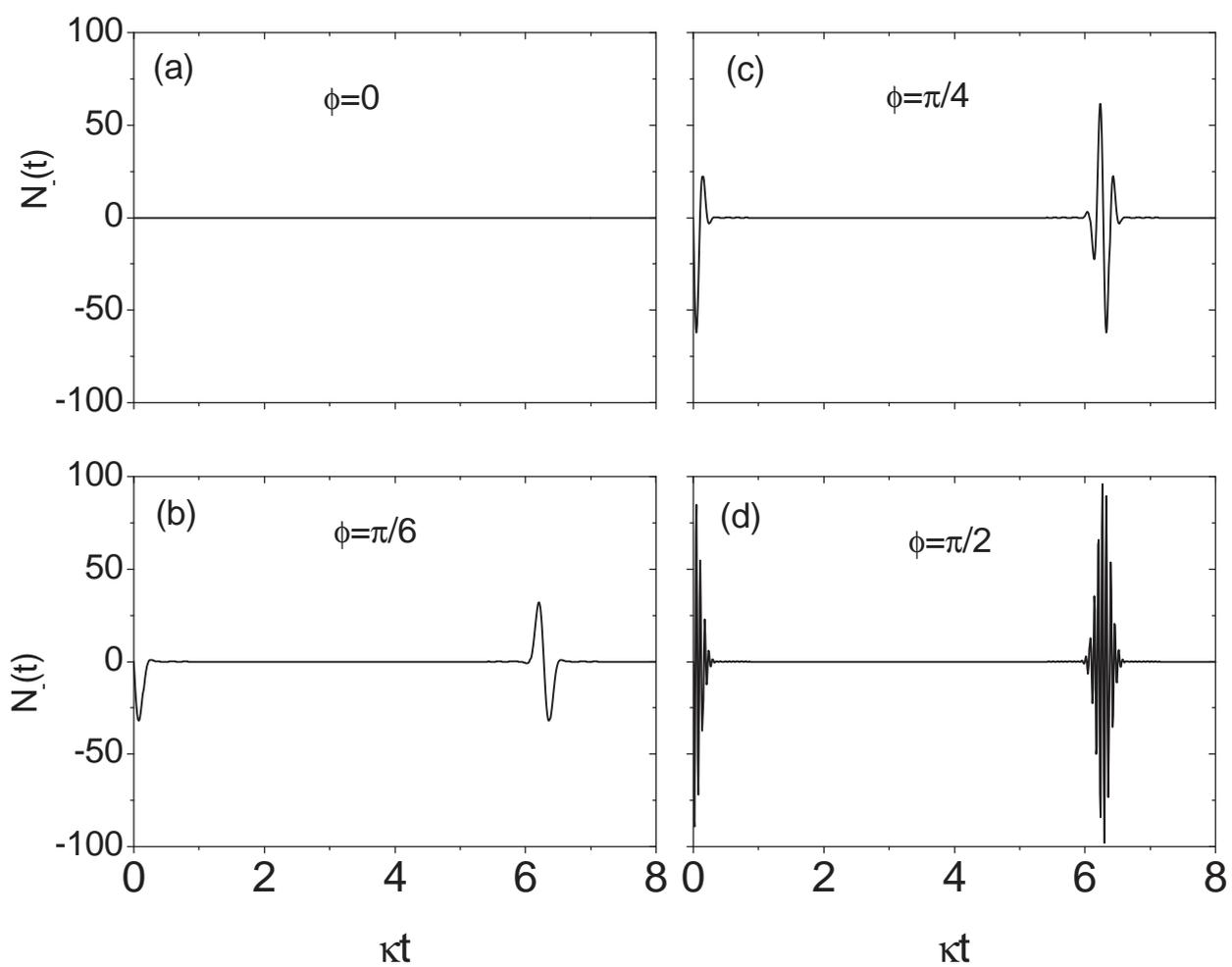
{\bf Fig. 4. }Population imbalance $N_{-}(t)$ for the value of relative
phase (a) $\phi =0$, (b) $\phi =\pi /6$, (c) $\phi =\pi /4$, and (d) $\phi
=\pi /2$. Other parameters are $N=100$ and $\kappa =0.01\Omega $.

\end{document}